\documentclass[aps,prd,showkeys,nofootinbib,superscriptaddress,preprint]{revtex4-1}
\usepackage{enumerate}
\usepackage{float}
\usepackage{amsmath,slashed,tensor,amssymb,epsfig,amsthm,bm,graphicx,graphics,slashed}
\usepackage[caption=false]{subfig}
\usepackage{hyperref}
\usepackage[utf8]{inputenc}
\usepackage[top=1.5cm, bottom=1.5cm, left=2cm, right=2cm]{geometry}

\usepackage{makecell}
\usepackage{microtype}
\usepackage[font=footnotesize,labelfont=bf]{caption}
\newcommand{\be}{\begin{equation}}
\newcommand{\ee}{\end{equation}}
\usepackage{color}
\definecolor{darkgreen}{rgb}{0,0.35,0}

\newcommand{\ucharlesipnp}{Institute of Particle and Nuclear Physics, Faculty of Mathematics and Physics,
Charles University, V Hole\v{s}ovi\v{c}k\'ach 2, 18000 Prague 8, Czech Republic}

\begin{document}

\title{Vortex solutions of Liouville equation and quasi spherical surfaces}

\author{Alfredo Iorio}
\email{alfredo.iorio@mff.cuni.cz}
\affiliation{\ucharlesipnp}

\author{Pavel K\r{u}s}
\email{pavel.kus.student@gmail.com}
\affiliation{\ucharlesipnp}


\begin{abstract}
We identify the two-dimensional surfaces corresponding to certain solutions of the Liouville equation of importance for mathematical physics, the non-topological Chern-Simons (or Jackiw-Pi) vortex solutions, characterized by an integer $N \ge 1$. Such surfaces, that we call $S^2 (N)$, have positive constant Gaussian curvature, $K$, but are spheres only when $N=1$. They have edges, and, for any fixed $K$, have maximal radius $c$ that we find here to be $c = N / \sqrt{K} $. If such surfaces are constructed in a laboratory by using graphene (or any other Dirac material), our findings could be of interest to realize table-top Dirac massless excitations on nontrivial backgrounds. We also briefly discuss the type of three-dimensional spacetimes obtained as the product $S^2 (N) \times \mathbb{R}$.
\end{abstract}


\maketitle

\section{Introduction}	

Liouville equation is an important equation of mathematical physics, that originally arose in the study of the local properties of two-dimensional surfaces \cite{Liouville},
\begin{equation} \label{LE:Liouville_equation_intro}
\Delta \ln \phi = - K \phi^2 \;,
\end{equation}
where $\Delta = \partial^2_{\tilde{x}} + \partial^2_{\tilde{y}}$, $\phi \equiv \phi(\tilde{x},\tilde{y})$, $K$ is the constant Gaussian curvature of the surface, and its infinitesimal line element, written in isothermal coordinates, $(\tilde{x},\tilde{y})$, is
\begin{equation}\label{isolineelement}
dl^2 = \phi^2 (d\tilde{x}^2 + d\tilde{y}^2) \,.
\end{equation}

\begin{figure}%
	\centering
	\includegraphics[width=5.5cm]{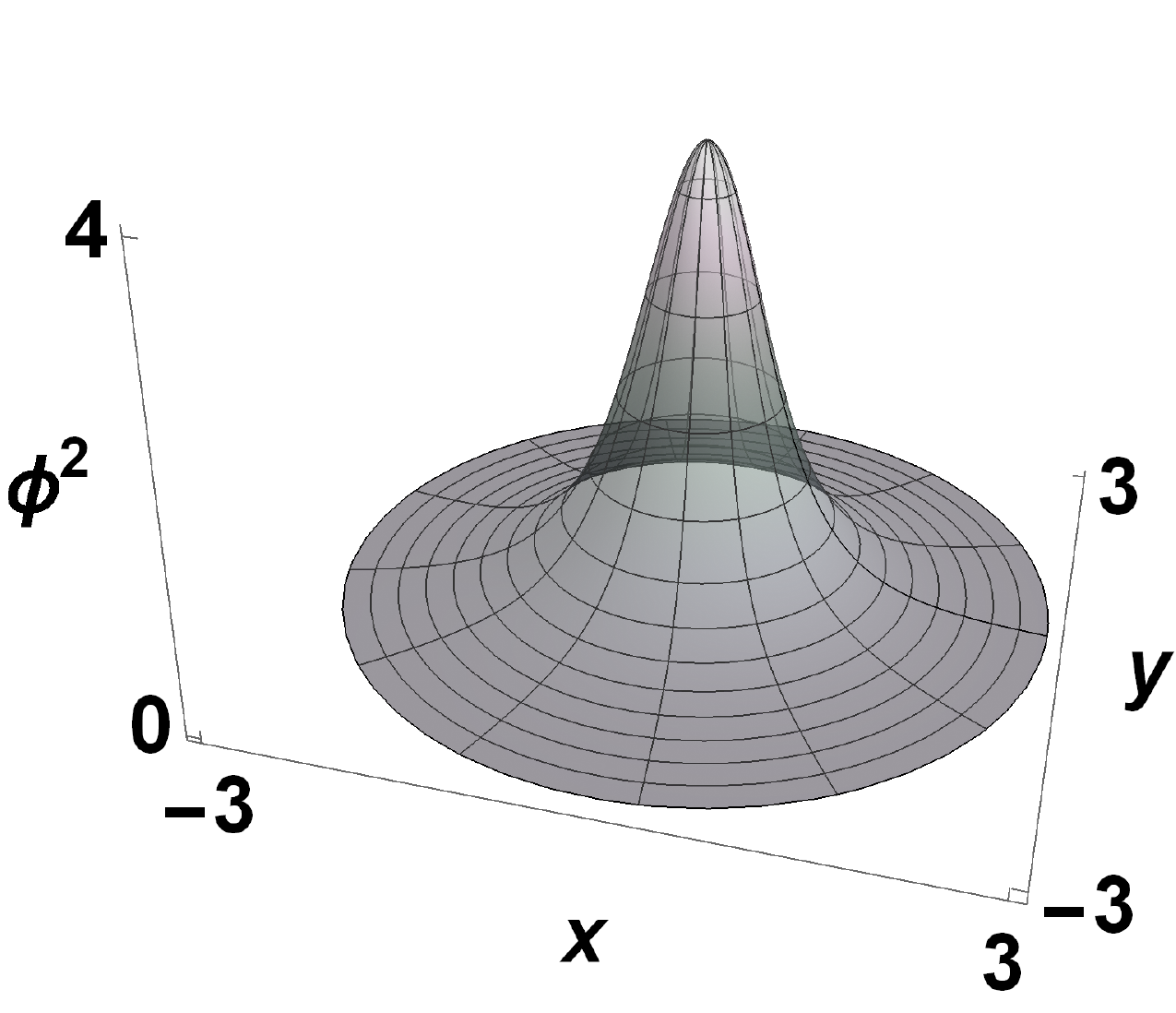}
	\includegraphics[width=5.5cm]{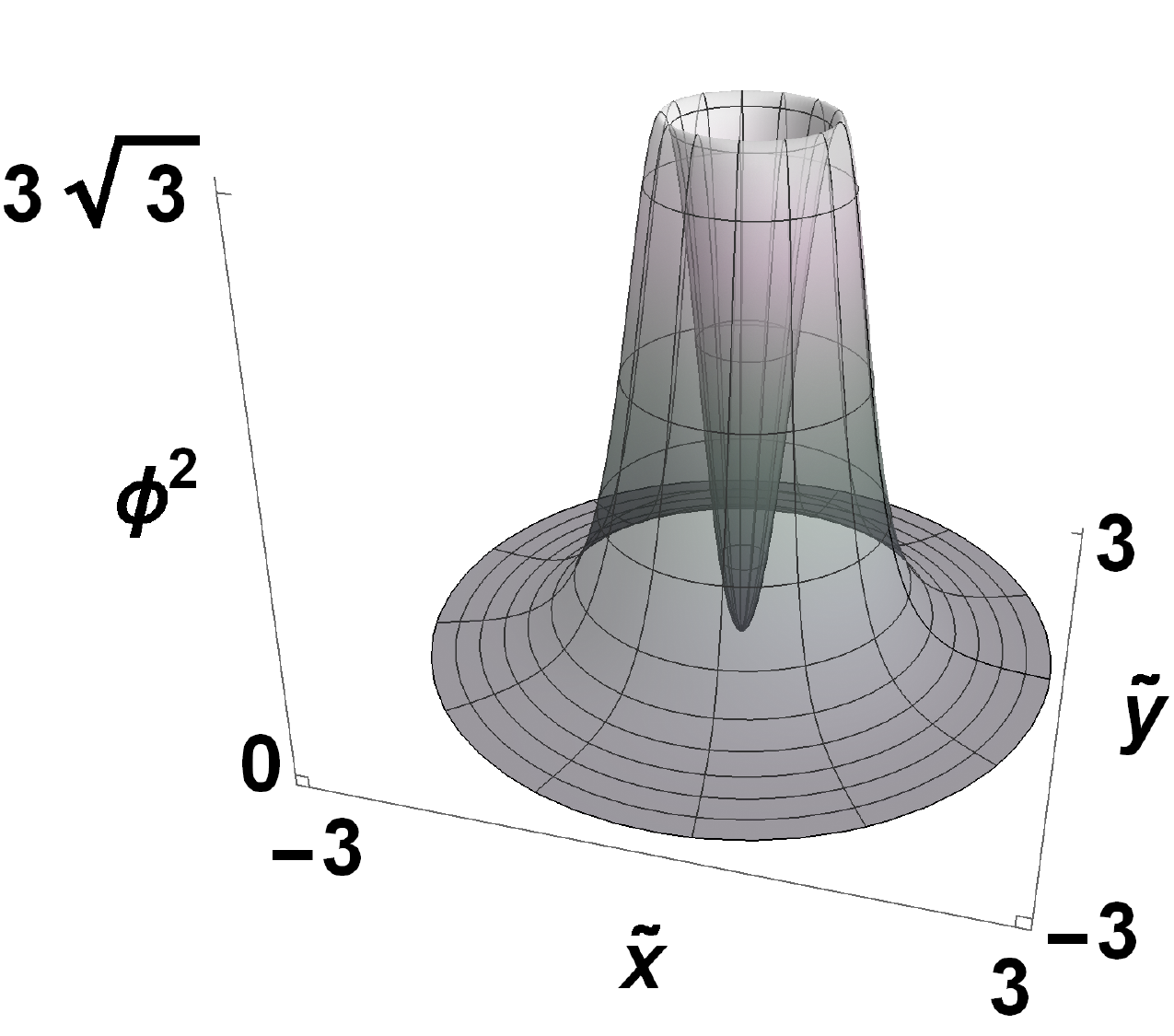}
	\includegraphics[width=5cm]{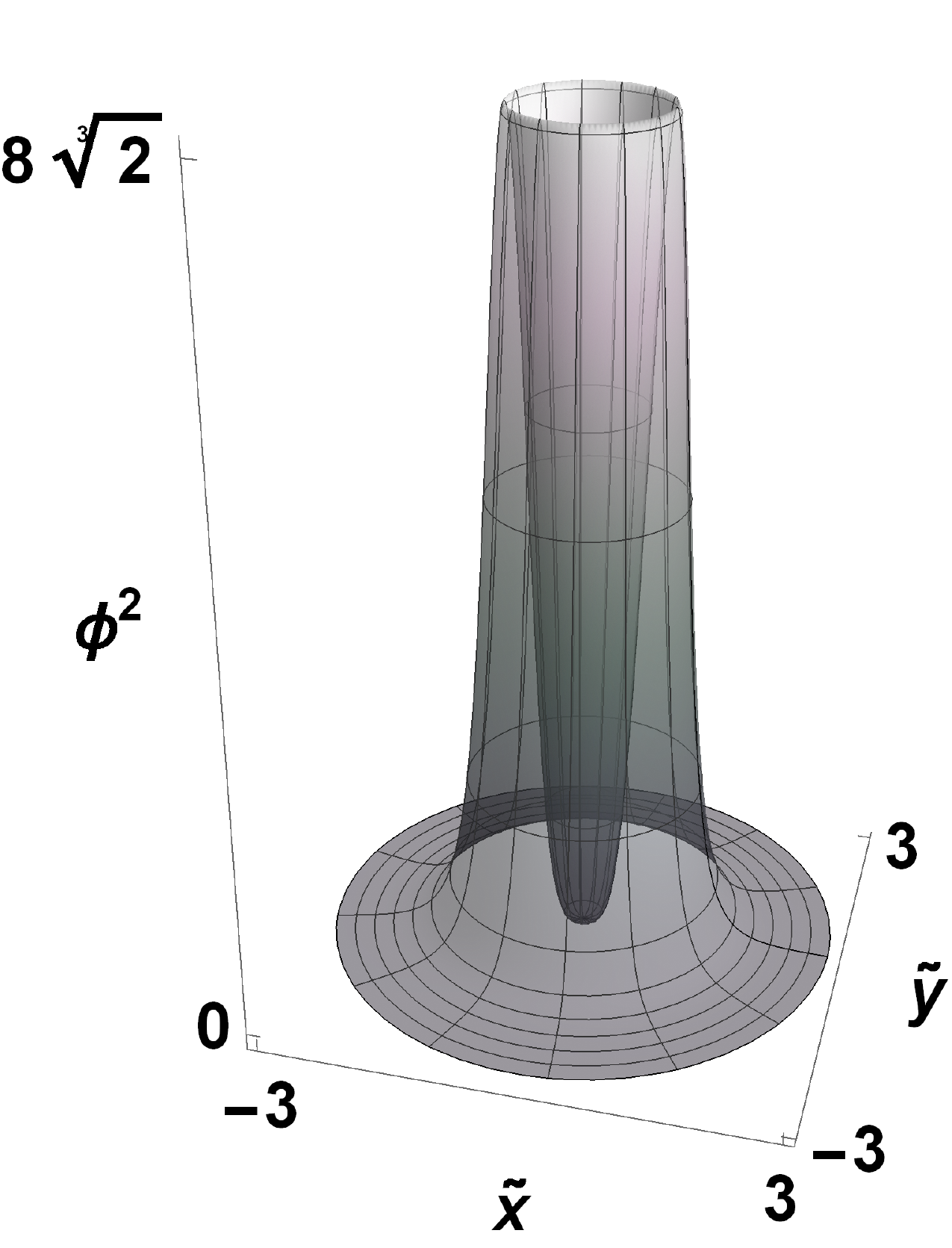}
	\caption{We plot here $\phi^2 (\tilde{r})$ for $K=1$, and $N=1$ (left), $N=2$ (centre), and $N=3$ (right). Of course, the range of $\tilde{r}$ is infinite, in general, but for illustrative purposes, we only considered $\tilde{r} \in [0,3]$.}%
	\label{Fig:introduction}
\end{figure}

All solutions were found by Liouville \cite{Liouville}
\begin{equation}\label{liouvillesolutions1}
\phi = \frac{2}{\sqrt{|{K}|}}\frac{ |f'(z)|}{1\pm |f(z)|^2},
\end{equation}
where '$+$' corresponds to $K>0$, while '$-$' corresponds to $K<0$, $z \equiv \tilde{x} + i \tilde{y}$ (notice that $\Delta_\mathrm{(\tilde{z},\bar{z})} = 4 \partial_{z}\partial_{\bar{z}}$), and $f$ is any meromorphic function, with at most simple poles, which satisfies $f'\equiv df/dz \neq 0$ for every $z$ in a simply connected domain.

Given its ties with exactly solvable two-dimensional models, this equation has many applications in disparate areas of the theoretical investigation, from string theory and low-dimensional gravity, to conformal field theories and condensed matter. One such area is Chern-Simons theory, for which Jackiw and Pi have found vortex solutions \cite{jackiwpi}, known as Jackiw-Pi (JP) vortices (for reviews see \cite{jackiwpiReview} and \cite{Horvathy_Zhang}). Indeed, such vortices arise as solutions of the self-duality equations that, in the non-topological case (that is, asymptotic to zero at infinity) lead to the Liouville equation \cite{Horvathy_Yera}. In that case
\begin{equation}\label{radialAnsatz}
f(z) = z^{-N}
\end{equation}
with $N \ge 1$ a natural number, and the curvature taken to be $K>0$.

The profiles of such vortices, studied in \cite{Horvathy_Yera} (see also \cite{Horvathy_Zhang}), have been obtained, and some are represented here in Fig.\ref{Fig:introduction}. Nonetheless, those are the profiles of the \textit{conformal factors} of the surface, not the profiles of \textit{corresponding surfaces}. The main goal of this paper is to find the latter.

Notice that, being $N=1,2, \cdots, \infty$, for any given $K>0$, one should expect an infinite number of such surfaces, but is it not the sphere, $S^2$, the only surface of constant $K>0$? We shall fully solve this puzzle in what follows.

Besides being an interesting (and challenging) thing to do in its own right, to solve this problem could be of help in practical cases, when the coordinate system is of physical relevance. One example that comes to the mind is when the two-dimensional surface is a graphene membrane, on which Dirac quasi-particles live. In fact, this paper solves a problem posed in \cite{Iorio_weyl_symmetry} (where such configurations emerged in the study of the Weyl symmetry of graphene) and faced in \cite{kus}.

In Section \ref{sec3:shapes} we recall the needed results of the differential geometry of surfaces. We then find, in Section \ref{sec5:surfaces_JP}, the appropriate transformation from the abstract isothermal to the Cartesian coordinates for the JP vortices, leaving to Appendix A an alternative transformation. In Section \ref{sec6:negative_K} we show that no straightforward generalization to negative constant Gaussian curvature can be found. In Section \ref{sec7:associated_spactimes} we briefly discuss spacetimes of the kind $S^2(N) \times \mathbb{R}$. In the last Section, we draw our conclusions, and report a useful identity in Appendix B.

\section{Surfaces of revolution of constant $K$}
\label{sec3:shapes}

Let us now briefly recall the results of the classic differential geometry of surfaces that we shall need. We shall refer mostly to \cite{eisenhart} and \cite{Spivak}, but see also \cite{Iorio_curved_spacetime}.

As we are dealing with radially symmetric solutions, let us focus on surfaces of revolution. A suitable parametrization in $\mathbb{R}^3$ is the canonical \cite{eisenhart}
\begin{equation}
\label{eq:surfaces_of_revolution_parametrization}
x(u,v) = R(u)\cos v,~
y(u,v) = R(u)\sin v,~
z(u) = \pm \int_{}^{u} \sqrt{1-\left[ R^{'}(\bar{u})\right]^2 }\mathrm{d}\bar{u},
\end{equation}
where $R(u)$ is a radial function which characterizes the given surface, $v\in[0,2\pi]$ is the longitude, $u \in [u_{\mathrm{min}}, u_{\mathrm{max}}]$ is the latitude, and $ R^{'} \equiv {dR}/{du}$. Notice that the range of the latitude, $u$, is dictated by the request $\left[ R^{'}(u)\right]^2 \le 1$.

 \begin{figure}%
	\centering
	\includegraphics[width=5.5cm]{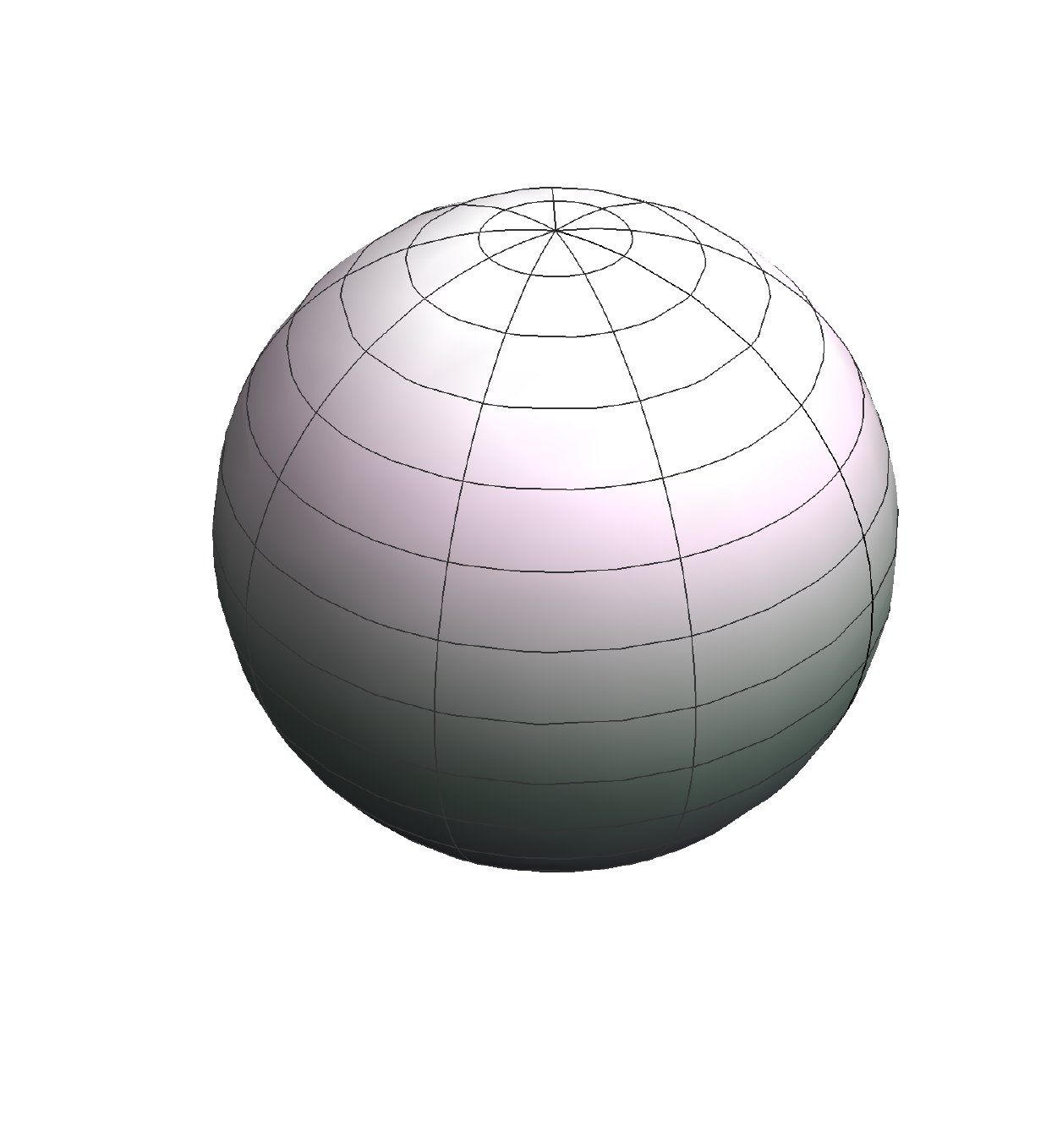}
	\includegraphics[width=6.0cm]{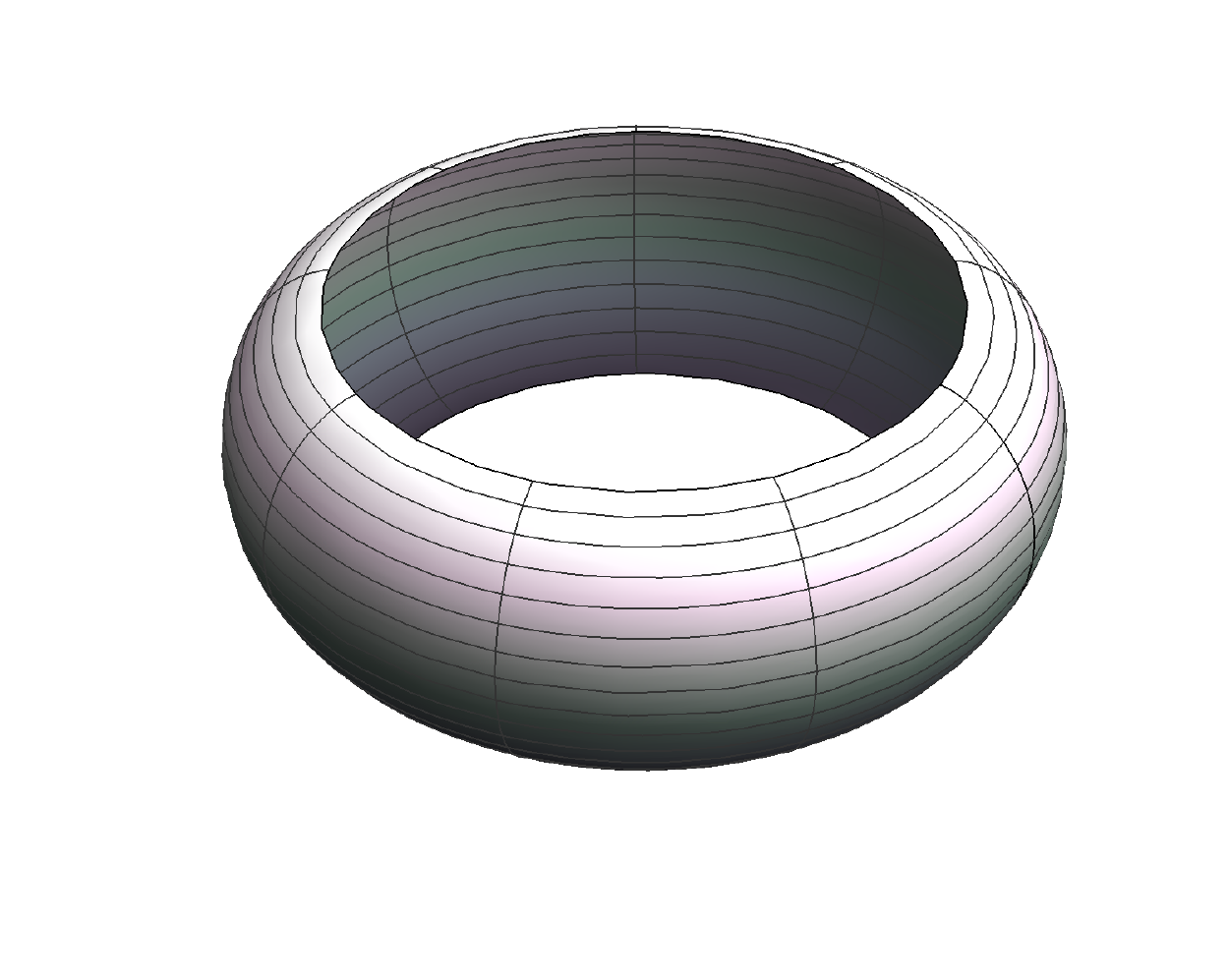}~~~~
	\includegraphics[width=3.5cm]{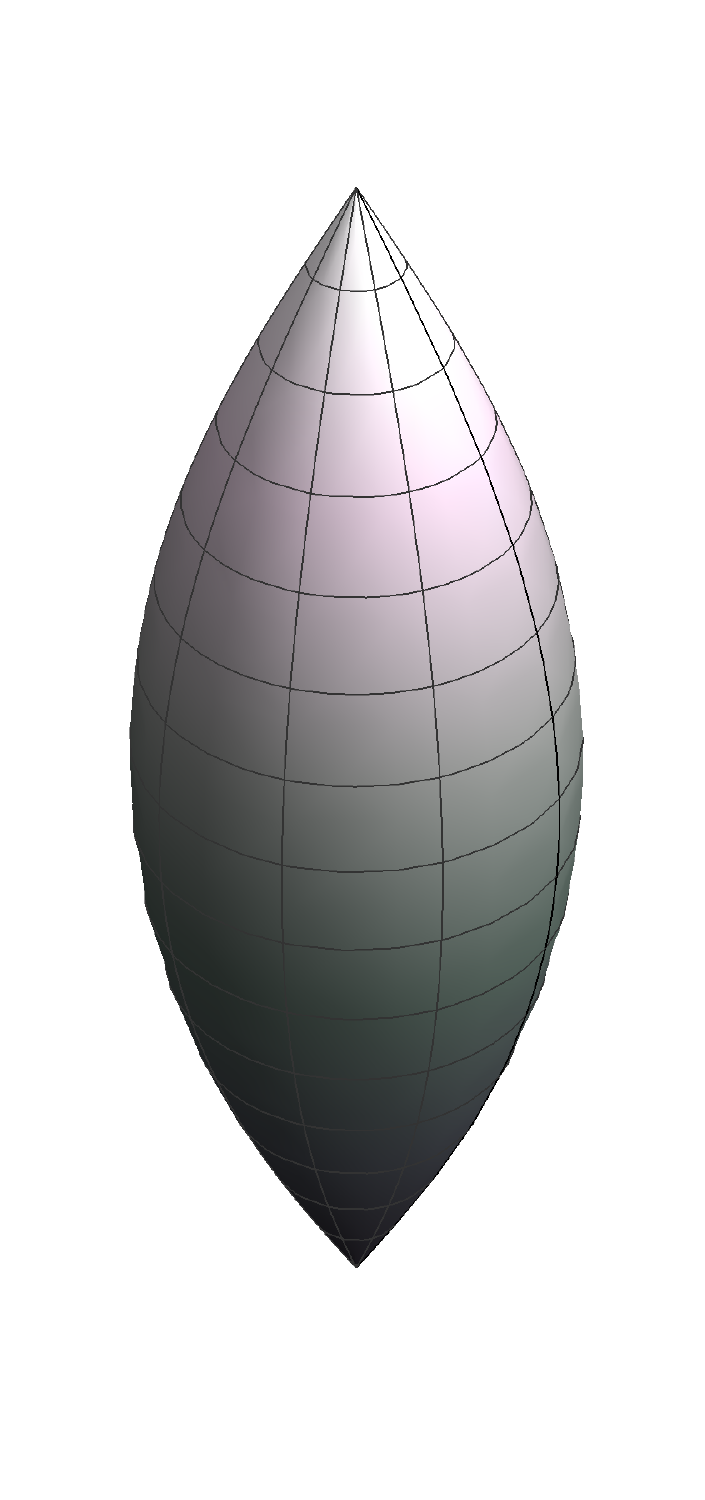}
	\caption{The three members of the sphere's family, left to right: sphere, bulge and spindle surfaces.}%
	\label{Fig:graphs_surfaces_of_revolution_with_constant_K positive}
\end{figure}

In canonical coordinates, the infinitesimal line element becomes
\begin{equation}
\label{Appendix:line_element_sphere}
dl^2 \equiv dx^2 + dy^2 + dz^2 \equiv du^2 + R^2(u)dv^2 \,,
\end{equation}
from which, one reads--off the relations between Gaussian curvature, $K$, and $R(u)$
\begin{equation}
\label{ode1}
K = -\frac{R^{''}(u)}{R(u)}\,,
\end{equation}
that is an ordinary differential equation for $R$. Say $K \equiv \pm 1/a^2 =$const, then (\ref{ode1}) has the following solutions
\begin{eqnarray}
R(u) & = & c\cos\left( \frac{u}{a} + b\right) \quad \quad \quad \quad \quad \mathrm{for} \, K = \frac{1}{a^2} \,, \label{Appendix:eq:ode_sol} \\
R(u) & = & c_1\sinh\frac{u}{a} + c_2\cosh\frac{u}{a} \quad \quad \mathrm{for} \,  K=-\frac{1}{a^2}\,, \label{ode_sol1}
\end{eqnarray}
where $a$ is a positive real number, and $b\,,c\,,c_1\,, c_2$ are integration constants. We can set the origin of $u$ such that $b = 0$.

When $K = 1/a^2 > 0$, there is more than the sphere. Depending on how $c$ and $a$ are related, we have
\begin{itemize}
	\item [1.] Sphere:  $c = a$\,, $u/a \in [- \pi/2, \pi/2]$\,,
	\item [2.] Bulge surface: $c > a$\,,  $u/a \in [- \arcsin\left( {a}/{c}\right), \arcsin\left( {a}/{c}\right)]$\,,
	\item [3.] Spindle surface: $c < a$\,,  ${u}/{a} \in [-{\pi}/{2}, {\pi}/{2}]$\,,
\end{itemize}
in all cases, $v\in [0, 2 \pi]$. See Fig. \ref{Fig:graphs_surfaces_of_revolution_with_constant_K positive}.

\begin{figure}%
	\centering
	\includegraphics[width=5cm]{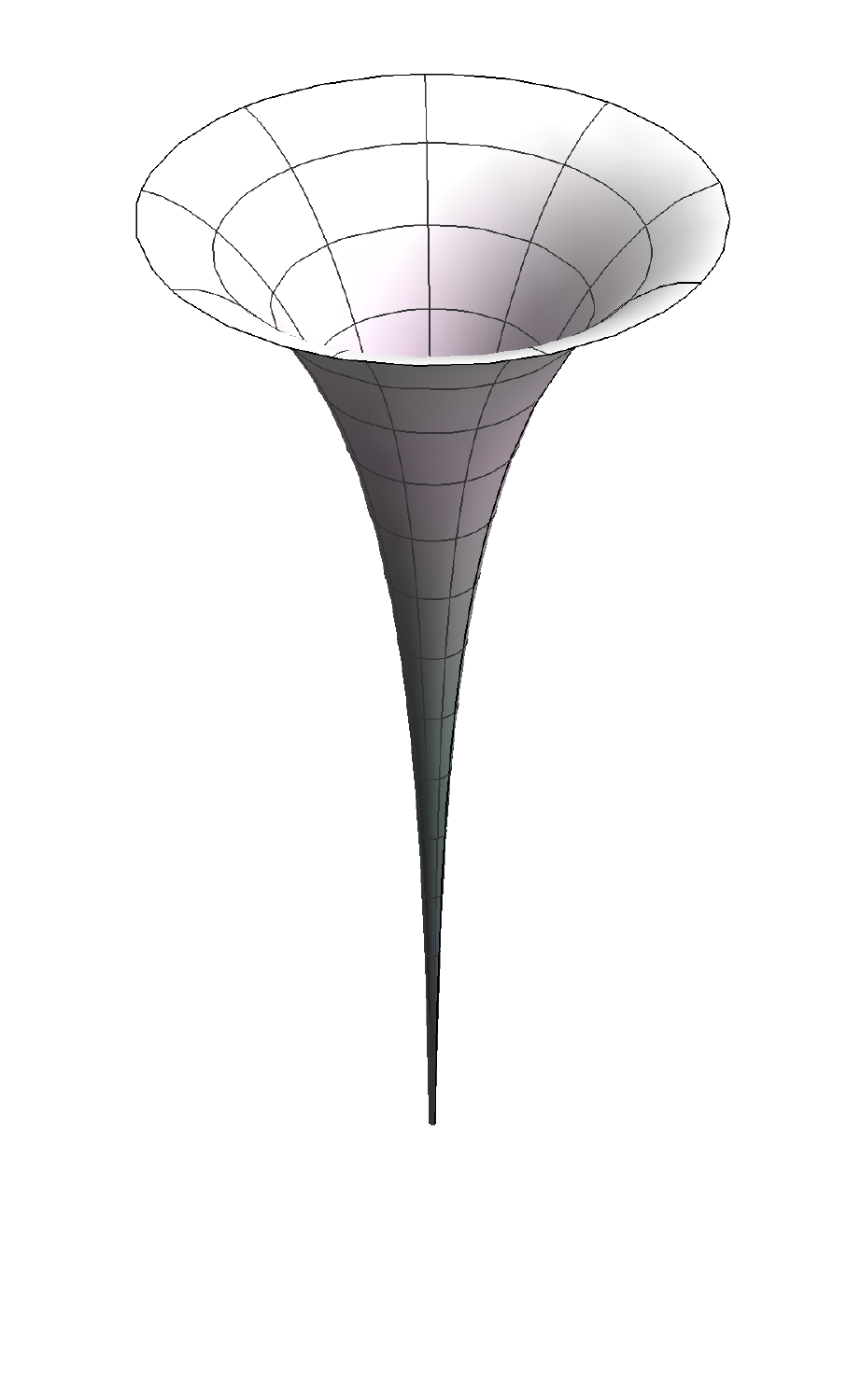}
	\includegraphics[width=6cm]{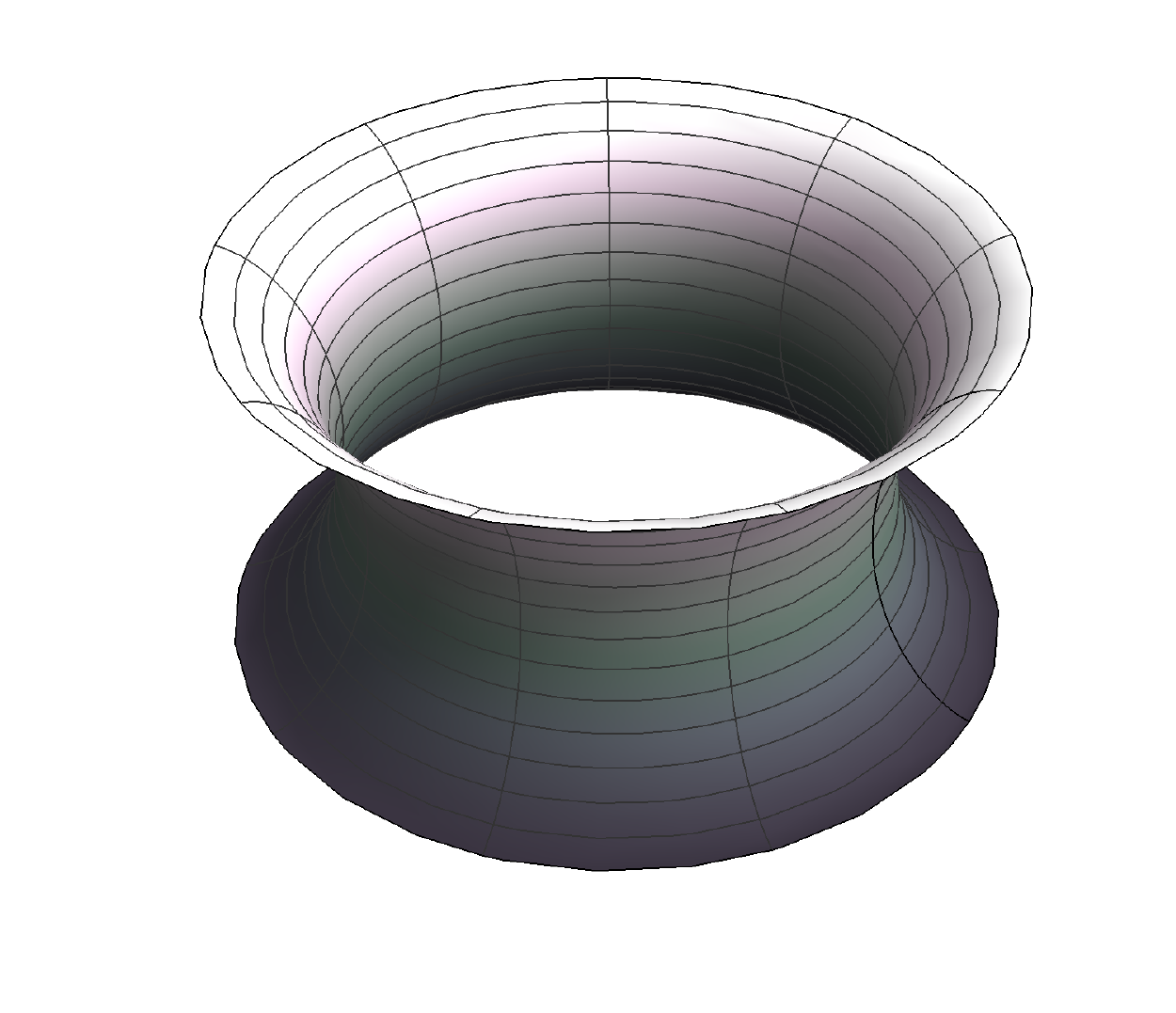}
	\includegraphics[width=6cm]{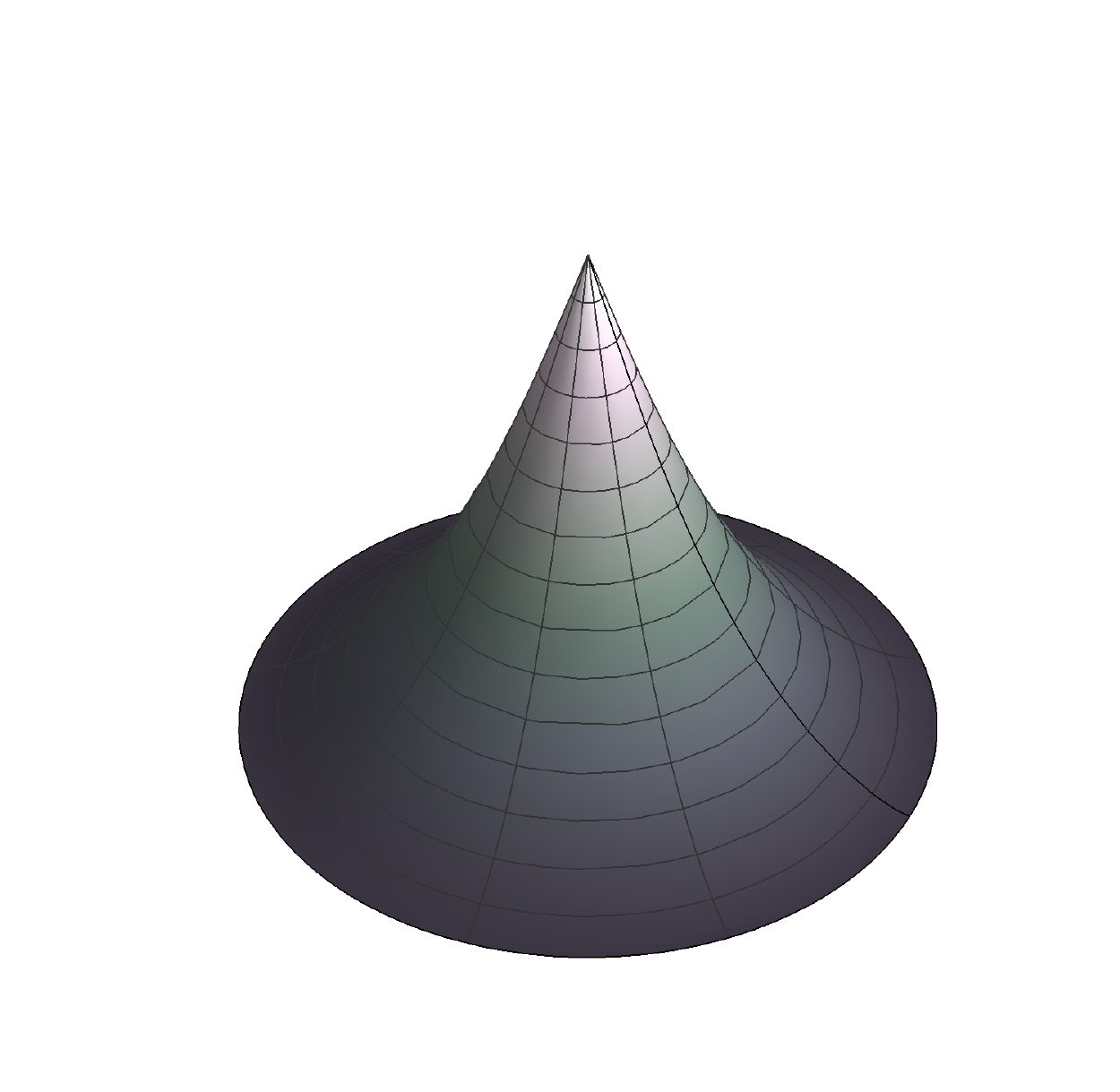}
	\caption{The three pseudospheres, left to right: Beltrami, hyperbolic and elliptic.}%
\label{Fig:graphs_surfaces_of_revolution_with_constant_K negative}
\end{figure}

When $K = -1/a^2 < 0$, we have the three pseudospheres
\begin{itemize}
	\item [1.] Beltrami:  $c_1 = c_2 \equiv c > 0$\,,  $u/a \in [-\infty, \ln({a}/{c})]$\,,
	\item [2.] Hyperbolic: $c_2 \equiv c$\,, $c_1 = 0$\,,
	
	 $u/a \in [-{\rm arccosh} \left( \sqrt{1+({a}/{c})^2}\right), {\rm arccosh} \left( \sqrt{1+({a}/{c})^2}\right)]$\,,
	\item [3.] Elliptic: $c_1 \equiv c$\,, $c_2 = 0$\,,  $u/a \in [0, {\rm arcsinh}\cot\beta]$ with $c \equiv a\sin\beta$\,,
\end{itemize}
in all cases, $v\in [0, 2 \pi]$. See Fig. \ref{Fig:graphs_surfaces_of_revolution_with_constant_K negative}.

The two families look pretty similar. In both, we have singularities: edges for bulge and hyperbolic (and Beltrami) surfaces, cusps for spindle and elliptic surfaces. Nonetheless, there is a crucial difference
between the families: for $K>0$ such singularities are removable by a simple redefinition of the longitude $v \longmapsto \bar{v} = ({c}/{a})v$, while there is no way to remove the singularities when $K<0$. The latter is the effect of a theorem proved by Hilbert, stating ``{\it There exists no analytical, complete surface of constant negative Gaussian curvature in Euclidean three space}'', see, e.g., \cite{ovchinnikov}.

Let us make here two comments. First, from the above it is clear why the other two members of the ``sphere's family'', the bulge and the spindle surfaces, did not share the same fortune of the sphere. If one can simply redefine an angle variable, and get rid of these obnoxious singularities, why not doing it? In the following, we shall give (to our knowledge, for the first time) a good reason not to do so.

The second comment is relative to the surfaces of negative constant $K$. Besides the three pseudospheres, there is actually an infinite number of such surfaces, possibly related to the infinite number of finite Fuchsian groups, that are the Lobachevsky version of the crystallographic groups of the Euclidean space, on this see \cite{mclachlan} and \cite{grapheneExperiments2}. It is natural to ask, then, whether the infinite numbers of vortices might be related to this infinite, where the positive curvature family only has one serious representative.

Thus our goal is to solve these puzzles, by finding the coordinate transformations relating the conformal factors to the actual surfaces. Let us write
\begin{equation}
\label{eq:conformal_factor_H-Y}
	\phi_{+}(\tilde{r}) = \frac{2N}{\sqrt{K}}\frac{\tilde{r}^{N-1}}{\tilde{r}^{2N} + 1},
\end{equation}
where we shall call $\phi_{+}$ ($\phi_{-}$) the solutions corresponding to $K>0$ ($K<0$), and we denote the isothermal radial coordinate as $\tilde{r} \equiv \sqrt{\tilde{x}^2 + \tilde{y}^2} \equiv |z|$.

The problem of finding the coordinate transformation, from the radial isothermal coordinates $(\tilde{r}, \tilde{\theta})$, to the Cartesian coordinates $(x,y,z)$, it is simple to state, but being the related system of partial differential equations (pds), that needs to be solved, nonlinear, the problem is, in general, a difficult one. Let us proceed.

\section{Surfaces corresponding to JP vortices}
\label{sec5:surfaces_JP}

We shall use radial coordinates for the surfaces, $(\tilde{r},\tilde{\theta})$, hence, considering the expression (\ref{eq:conformal_factor_H-Y}) above, the line element (\ref{isolineelement}), in these coordinates, is \begin{equation}
\label{eq:line_element_generally}
dl^2 = \phi_{+}^2(\tilde{r})(d\tilde{r}^2+\tilde{r}^2d\tilde{\theta}^2) \,.
\end{equation}

Consider now the surface imbedded in $\mathbb{R}^3$, hence $dl^2 = dx^2 +dy^2 +dz^2$, must be rewritten with $x(\tilde{r},\tilde{\theta}), y(\tilde{r},\tilde{\theta}),z(\tilde{r},\tilde{\theta})$. The result is
\begin{eqnarray}
dl^2 & = & \left[\left( \frac{\partial x}{\partial \tilde{r}}\right)^2 + \left( \frac{\partial y}{\partial \tilde{r}}\right)^2 + \left( \frac{\partial z}{\partial \tilde{r}}\right)^2 \right]d\tilde{r}^2 + \left[\left( \frac{\partial x}{\partial \tilde{\theta}}\right)^2 + \left( \frac{\partial y}{\partial \tilde{\theta}}\right)^2 + \left( \frac{\partial z}{\partial \tilde{\theta}}\right)^2 \right]d\tilde{\theta}^2
\nonumber \\
& + & 2\left[\frac{\partial x}{\partial \tilde{r}} \frac{\partial x}{\partial \tilde{\theta}} + \frac{\partial y}{\partial \tilde{r}} \frac{\partial y}{\partial d\tilde{\theta}} + \frac{\partial z}{\partial \tilde{r}} \frac{\partial z}{\partial \tilde{\theta}}\right]d\tilde{r}d\tilde{\theta}
\label{eq:line_element2} \,.
\end{eqnarray}
Comparing term by term the latter with (\ref{eq:line_element_generally}), leads to the announced system of non-linear pds
\begin{eqnarray}
\left( \frac{\partial x}{\partial \tilde{r}}\right)^2 + \left( \frac{\partial y}{\partial \tilde{r}}\right)^2 + \left( \frac{\partial z}{\partial \tilde{r}}\right)^2
& = & \phi_{+}^2(\tilde{r}) \,, \label{eq:coor_transf_1} \\
 \left( \frac{\partial x}{\partial \tilde{\theta}}\right)^2 + \left( \frac{\partial y}{\partial \tilde{\theta}}\right)^2 + \left( \frac{\partial z}{\partial \tilde{\theta}}\right)^2 & = & \phi_{+}^2(\tilde{r})\tilde{r}^2 \,,
\label{eq:coor_transf_2} \\
\frac{\partial x}{\partial \tilde{r}} \frac{\partial x}{\partial \tilde{\theta}} + \frac{\partial y}{\partial \tilde{r}} \frac{\partial y}{\partial \tilde{\theta}} + \frac{\partial z}{\partial \tilde{r}} \frac{\partial z}{\partial \tilde{\theta}} & = & 0 \label{eq:coor_transf_3} \,.
\end{eqnarray}
This is clearly a difficult problem. Let us try with\footnote{A similar Ansatz, but with a different role of $N$, is proposed in the Appendix.}
\begin{equation}
\label{VS:eq:space_coordinates}
x = R_{+}(\tilde{r})\cos\tilde{\theta}, ~y = R_{+}(\tilde{r})\sin\tilde{\theta}, ~z = z_{+}(\tilde{r}) \,,
\end{equation}
where the range of $\tilde{\theta}$ is $[0,2\pi]$, for all $N$s.

Eq. (\ref{eq:coor_transf_3}) is satisfied immediately, while (\ref{eq:coor_transf_2}) only holds when
\begin{equation}
\label{VS:eq:radial_part}
R_{+}(\tilde{r}) =\tilde{r} \phi_{+}(\tilde{r}) = \frac{2N}{\sqrt{{K}}}\frac{\tilde{r}^{N}}{ \tilde{r}^{2N}+1} \,,
\end{equation}
that is compatible with (\ref{eq:conformal_factor_H-Y}). For the $z$ coordinate holds
\begin{equation}
\label{eq:z_coordinate}
z_{+} = \int \sqrt{\phi_{+}^2(\tilde{r}) - [R_{+}'(\tilde{r})]^2} d\tilde{r}.
\end{equation}
To have $z$ real, the integrand must be nonnegative
\begin{equation}
\frac{4N^2}{K}\frac{\tilde{r}^{2N-2}}{(\tilde{r}^{2N} + 1)^2} - \frac{4N^4}{K}\left[ \frac{\tilde{r}^{N-1}(1-\tilde{r}^{2N})}{(\tilde{r}^{2N} + 1)^2}\right]^2\ge0.
\end{equation}
We obtain\footnote{For $N = 1$, corresponding to the full sphere, we shall take $\tilde{r}_{\mathrm{min}} = 0$ and $\tilde{r}_{\mathrm{max}} = +\infty$. That is, we shall consider not to be a problem that $\tilde{r}_{\mathrm{max}}$, for $N=1$, includes a division by zero. We shall work with these relations for all $N$s.}, $\tilde r_{\mathrm{min}}$ and $\tilde r_{\mathrm{max}}$
\begin{equation}
\label{eq:r_min_r_max}
\tilde{r}_{\mathrm{min}} = \sqrt[{2N}]{\frac{N-1}{N+1}},~~ \tilde{r}_{\mathrm{max}} = \sqrt[{2N}]{\frac{N+1}{N-1}}.
\end{equation}
This is a first interesting result: the plots of $\phi^2_{+}(\tilde{r})$ for $N\ge2$, when thought as related to a real surface, cannot be plotted over the whole range of $\tilde{r}$, but should be plotted only from $\tilde{r}_{\mathrm{min}}$ to $\tilde{r}_{\mathrm{max}}$, something we could not tell without this analysis.

We do have an analytical expression for the $z$ coordinate (\ref{eq:z_coordinate}), but it would add no valuable information to present its nasty expression here. In fact, to plot the profile of the surface we better find the transformation formulae from $(\tilde{r},\tilde{\theta})$ to $(u,v)$, where the $z$ coordinate looks simple, and it is much easier to do general considerations for all surfaces for different $N$s. Let us do that.

First, using (\ref{eq:conformal_factor_H-Y}) in (\ref{eq:line_element_generally}), and defining $\tilde{R}\equiv\ln \tilde{r}^{N}$, we can write
\begin{equation}
\label{eq:geometry_of_the_surfaces_1}
dl^2 = \frac{4N^2}{{K}}\frac{\tilde{r}^{2(N-1)}}{(\tilde{r}^{2N}+1)^2}(d\tilde{r}^2+\tilde{r}^2d\tilde{\theta}^2)=\frac{4}{K}\frac{e^{2\tilde{R}}}{(e^{2\tilde{R}} + 1)^2}d\tilde{R}^2 +\frac{4N^2}{K}\frac{e^{2\tilde{R}}}{(e^{2\tilde{R}} + 1)^2}d\tilde{\theta}^2 \,.
\end{equation}
With the substitution ${\mathcal{R}}\equiv \arctan e^{\tilde{R}}$, (and recalling that ${K} \equiv {1}/{a^2}$), we have
\begin{eqnarray}
dl^2 & = & 4a^2d\mathcal{R}^2 +  \frac{4a^2N^2\tan^2\mathcal{R}}{\left(\tan^2\mathcal{R} + 1 \right)^2}d\tilde{\theta}^2 =
4a^2d\mathcal{R}^2 + a^2N^2\sin^2(2\mathcal{R})d\tilde{\theta}^2 \nonumber \\
& = & d\tilde u^2 + a^2N^2\sin^2\frac{\tilde u}{a} d\tilde{\theta}^2 \label{eq:geometry_of_the_surfaces_2} \,,
\end{eqnarray}
where we used $\tilde u\equiv 2a \mathcal{R}$ in the last equality. Now, we may perform a shift, $\tilde u\rightarrow \tilde u - \pi/2$, which gives
\begin{equation}
\label{eq:line_element_result}
	dl^2 = d\tilde u^2 + a^2N^2\cos^2\frac{\tilde u}{a} d\tilde{\theta}^2 \,.
\end{equation}
This is exactly the line element of a surface of revolution, with
\begin{equation}\label{RandcN}
  R(\tilde u) = (a N) \cos(\tilde u/a) \,
\end{equation}
which gives
\begin{equation}
\label{eq:geometrical_sense_of_N}
	c = aN.
\end{equation}
Notice that the set of coordinates $(\tilde{u}, \tilde{\theta})$ and $(u,v)$ are the same.

Equation (\ref{eq:geometrical_sense_of_N}) is a main part of the result we were looking for. It makes clear the geometrical meaning of $N$, as a factor that scales the radius. We immediately use that to identify the surfaces: when $N=1$, it is $c=a$, hence we are referring to a sphere, while for $N\ge 2$, it is $c > a$, hence we are referring to the bulge surfaces.

However, we still miss one crucial step, for a full identification of JP vortices with bulge surfaces. We do not know the range of $u$. It could happen that we only can associate portions of spheres/bulge surfaces.

Let us recall the relation
\begin{equation}
u/a = 2\arctan\left( \tilde{r}^N\right) - \pi/2	
\end{equation}
 and insert (\ref{eq:r_min_r_max}) into it
\begin{equation}
(u/a)_{\mathrm{min}} = 2\arctan\left( \sqrt{\frac{N-1}{N+1}}\right) - \frac{\pi}{2} ,~~ (u/a)_{\mathrm{max}} = 2\arctan\left( \sqrt{\frac{N+1}{N-1}}\right) - \frac{\pi}{2}.
\end{equation}
In Appendix B we prove that, for any real positive number $p$, the following identity holds
\begin{equation}
\label{eq:useful_identity}
2 \arctan\sqrt{p} = \arcsin\left( \frac{p-1}{p+1}\right) +\frac{\pi}{2}.
\end{equation}
This implies
\begin{equation}
\label{VS:range_min}
(u/a)_{\mathrm{min}} = 2	\arctan \sqrt{\frac{N-1}{N+1}} - \pi/2 = - \arcsin(1/N),
\end{equation}	
\begin{equation}
\label{VS:range_max}
(u/a)_{\mathrm{max}} = 2	\arctan \sqrt{\frac{N+1}{N-1}} - \pi/2 =  +\arcsin(1/N).
\end{equation}

This is a pleasant result. For $N\ge 2$, this range precisely coincides with the range of the bulge surfaces. Just compare the above with ${u}/{a} \in [-\arcsin\left( {a}/{c}\right) , \arcsin\left( {a}/{c}\right)]$, from Section \ref{sec3:shapes}. Of course, for $N = 1$, we have ${u}/{a} \in [-\pi/2 , \pi/2]$, which corresponds to the sphere of radius $a$.

\begin{figure}%
 	\flushleft
	\includegraphics[width=8.0cm]{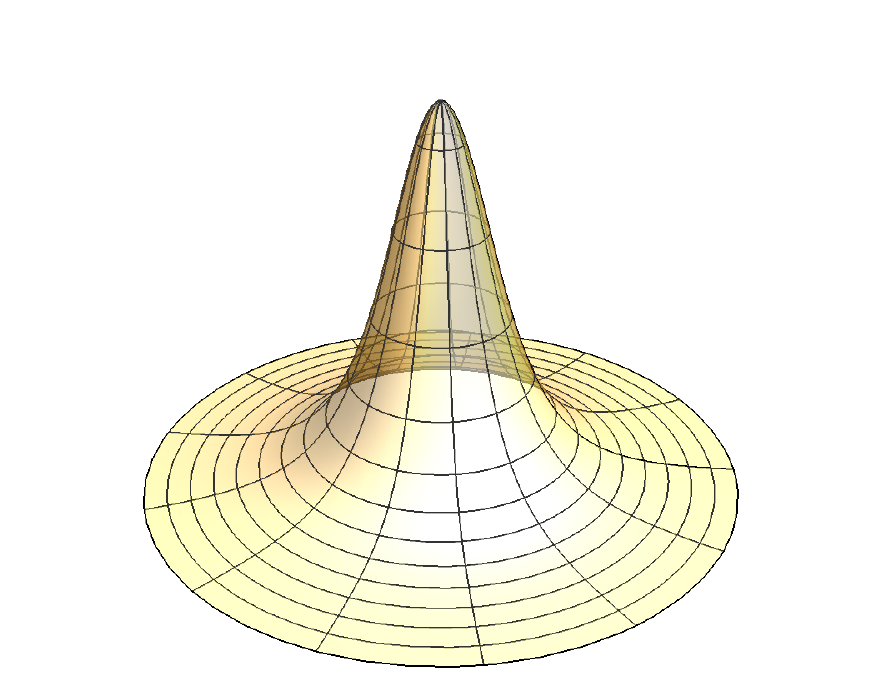} \hspace{1cm}
	\includegraphics[width=7.0cm]{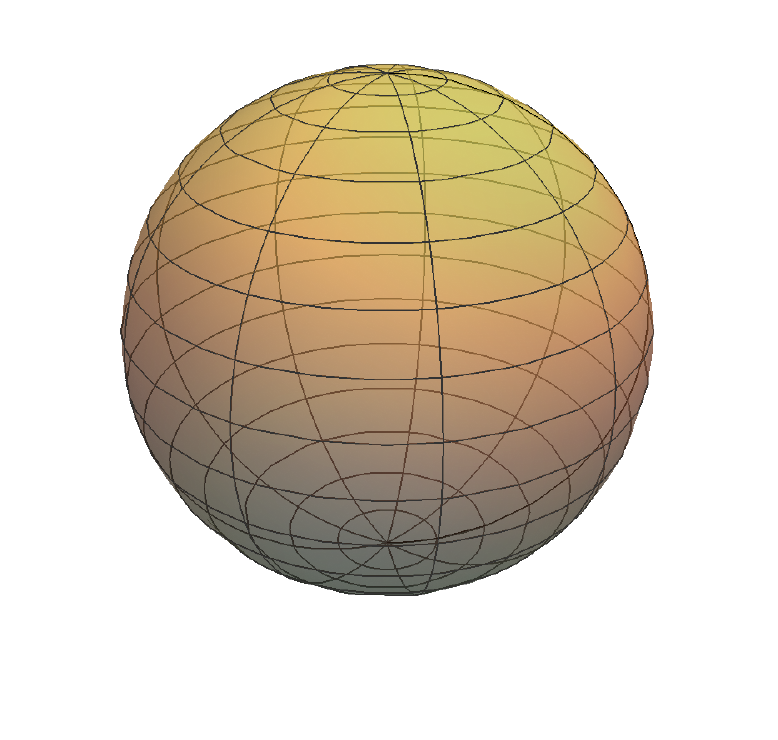}
	\includegraphics[width=7cm]{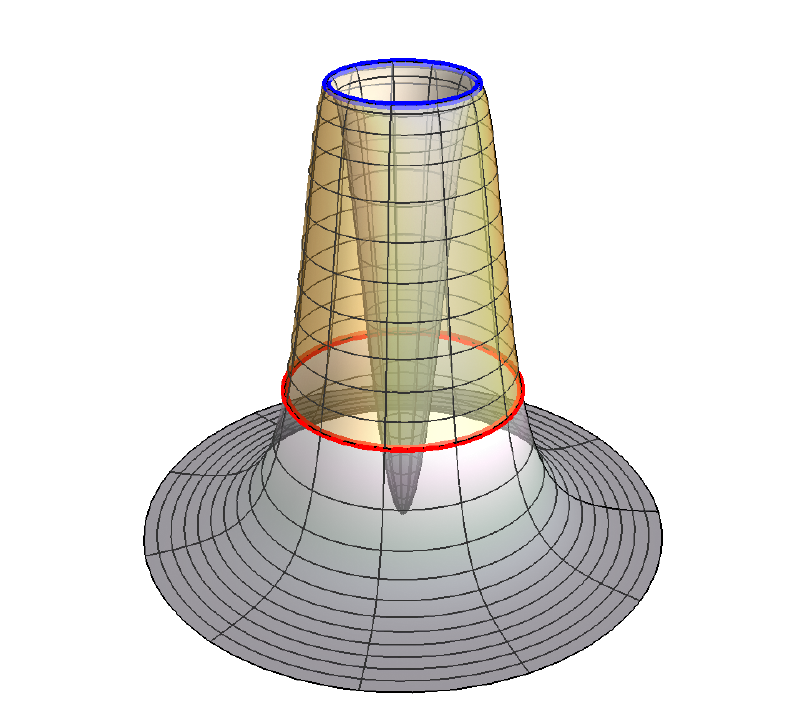} \hspace{1cm}
	\includegraphics[width=9cm]{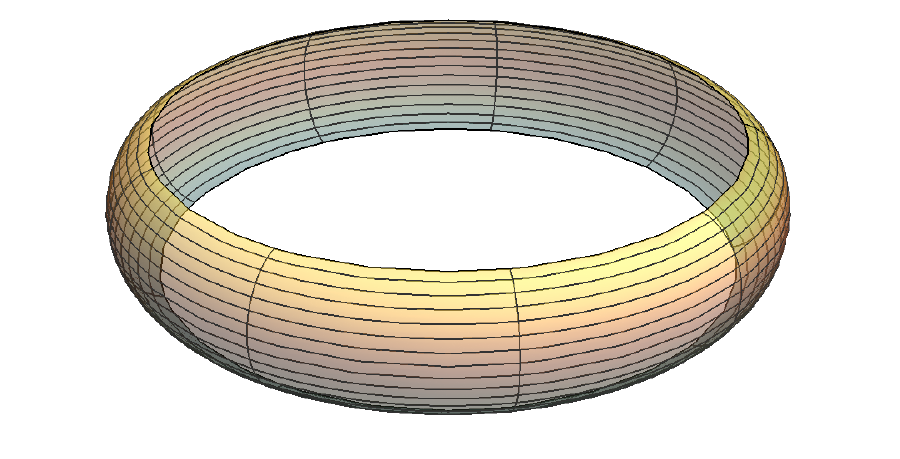}
	\includegraphics[width=6cm]{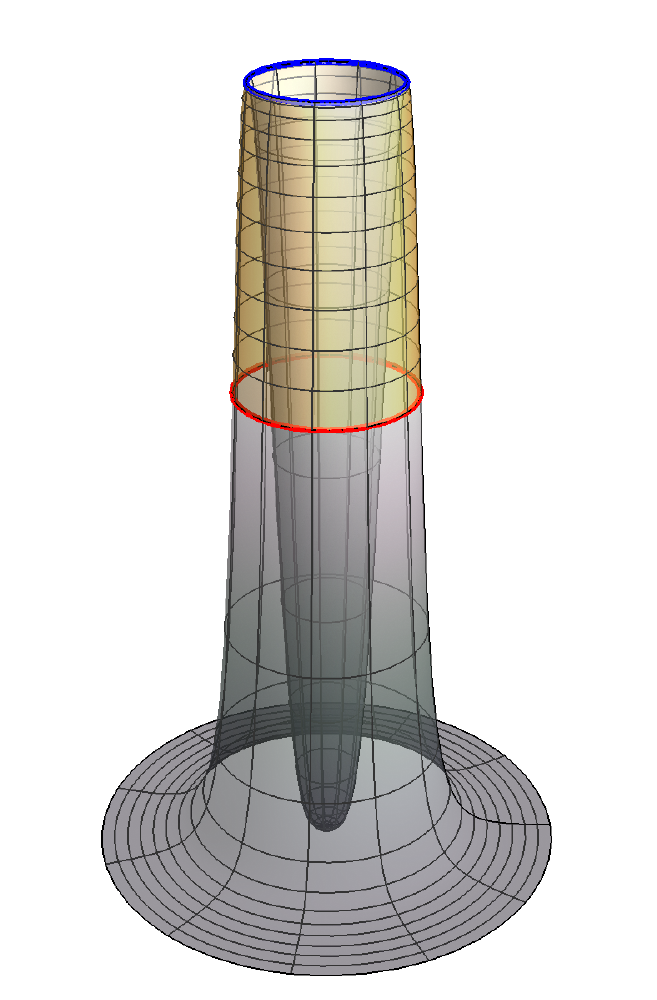} \hspace{1cm}
	\includegraphics[width=10cm]{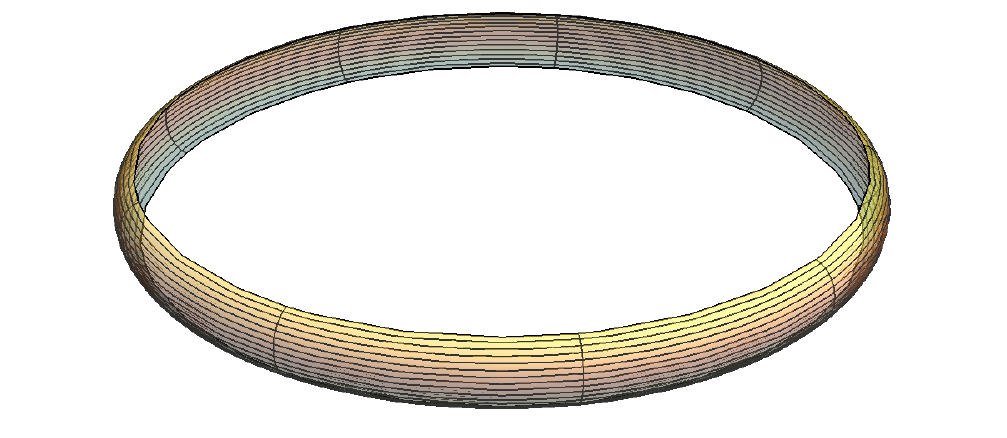}
	\caption{Here we depict our main result for three representative cases. That is, we plot the bulge surfaces next to the corresponding conformal $\phi^2$ , for $N=1$ (top), $N=2$ (center), and $N=3$ (bottom). In all cases, $a=1$, while $c = a N$. The highlighted regions on $\phi^2$ correspond to the permitted range, $(\tilde{r}_{\mathrm{min}}, \tilde{r}_{\mathrm{max}})$, (blue circle, red circle, respectively), according to the structure of the related surface, as established in (\ref{eq:r_min_r_max}).}%
	\label{Fig:graphs_surfaces_and_conformal_factors3}
\end{figure}

Let us call these surfaces, $S^2(N)$, with $S^2(1) = S^2$. Figures of selected surfaces and corresponding parts of conformal factors, plotted for $a=1$ and $N = 1,2,3$, are on Fig. \ref{Fig:graphs_surfaces_and_conformal_factors3}.
\newpage

\newpage
\clearpage
Let us close this Section by showing how (\ref{eq:z_coordinate}) reduces to  (\ref{eq:surfaces_of_revolution_parametrization}). It is easy to convince oneself that
\begin{equation}
	\frac{du}{d\tilde{r}} = \phi(\tilde{r}) \,,
\end{equation}
therefore (\ref{eq:coor_transf_1}) changes to
\begin{equation}
1 = \left( \frac{dR_+}{d\tilde{r}}\right)^2\left( \frac{d\tilde{r}}{du}\right)^2 + \left( \frac{d z}{d \tilde{r}}\right)^2\left( \frac{d\tilde{r}}{du}\right)^2 \,.
\end{equation}
After using the chain rule and integration of z coordinate, we obtain (\ref{eq:surfaces_of_revolution_parametrization}).

\section{JP vortices and negative Gaussian curvature}
\label{sec6:negative_K}

Here we would like to explore the possibility to extend the earlier study, to constant negative Gaussian curvature.

To start, we rewrite (\ref{eq:coor_transf_1})-(\ref{eq:coor_transf_3}) for $\phi_{+} \rightarrow \phi_{-}$, where
\begin{equation}
\label{eq:conformal_factor_minus}
	\phi_{-}(\tilde{r}) = \frac{2N}{\sqrt{|{K}|}}\frac{\tilde{r}^{N-1}}{\tilde{r}^{2N} - 1},
\end{equation}
with $K \equiv -1/a^2=$ const, and we make an Ansatz similar to the one made for $\phi_{+}$
\begin{equation}
\label{eq:space_coordinates_minus}
x = R_{-}(\tilde{r})\cos\tilde{\theta}, ~y = R_{-}(\tilde{r})\sin\tilde{\theta}, ~z = z_{-}(\tilde{r}) \,.
\end{equation}
The question is the same as for $\phi_{+}$: for what $\tilde{r}$ is the $z$ coordinate
\begin{equation}
\label{eq:z_coordinate_minus}
z_{-} = \int \sqrt{\phi_{-}^2(\tilde{r}) - [R_{-}'(\tilde{r})]^2} d\tilde{r}
\end{equation}
real? In other words, we must have $\phi_{-}^2(\tilde{r}) - [R_{-}'(\tilde{r})]^2\ge 0$. In fact,
\begin{equation}
\phi_{-}^2(\tilde{r}) - [R_{-}'(\tilde{r})]^2 \le
\frac{4N^2(1-N^2)}{|{K}|}\frac{\tilde{r}^{2N-2}}{(\tilde{r}^{2N} - 1)^2} \le 0 \,.
\end{equation}
Therefore, with the exception of the singularity at $\tilde{r} = 1$, where it is not defined, the integrand is always non-positive. This approach does not work.

Let us consider, a different option, i.e., to follow steps similar to those used for (\ref{eq:geometry_of_the_surfaces_1}) and (\ref{eq:geometry_of_the_surfaces_2}). The infinitesimal line element becomes
\begin{equation}
dl^2 = \frac{4N^2}{|K|}\frac{\tilde{r}^{2(N-1)}}{(\tilde{r}^{2N}-1)^2}(d\tilde{r}^2+\tilde{r}^2d\tilde{\theta}^2) = \frac{4}{|K|}\frac{e^{2\tilde{R}}}{(e^{2\tilde{R}} - 1)^2}d\tilde{R}^2 +  \frac{4N^2}{|K|}\frac{e^{2\tilde{R}}}{(e^{2\tilde{R}} - 1)^2}d\tilde{\theta}^2 ,
\end{equation}
where we used $\tilde{R}\equiv\ln \tilde{r}^{N}$ in the last equality. We can then substitute $\mathcal{R}~\equiv~\int \frac{e^{\tilde{R}}}{(e^{2\tilde{R}} - 1)}d\tilde{R} = {\rm arctanh} \, e^{\tilde{R}}$. The integration makes sense for $\tilde{R} \in (-\infty, 0)$, or for $\tilde{R} \in (0, +\infty)$, to avoid the singular point. However, because of the domain of the arctangent, we only consider the range $\tilde{R} \in (-\infty, 0)$, so that $\mathcal{R} \in [0,1]$. The line element becomes
\begin{equation}
	dl^2 = 4a^2d\mathcal{R}^2 + \frac{4a^2N^2\tanh^2\mathcal{R}}{\left(\tanh^2\mathcal{R} - 1 \right)^2}d\tilde{\theta}^2 = 4a^2d\mathcal{R}^2 + a^2N^2\sinh^2(2\mathcal{R})d\tilde{\theta}^2 \,,
\end{equation}
and one last substition, $u\equiv 2a \mathcal{R}$, leads to
\begin{equation}
	dl^2= du^2 + a^2N^2\sinh^2\frac{u}{a}d\tilde{\theta}^2 \,.
\end{equation}
This result is remarkably similar to the line element of the elliptic pseudosphere, presented in Section \ref{sec3:shapes}. However, for the elliptic pseudosphere the parameter $c$ is always strictly less than the radius of curvature, $a$, i.e., $c < a$, but here $c \equiv a N \ge a$.

We conclude that, although line elements similar to those of the elliptic pseudospheres naturally appear, those surfaces cannot be the negative curvature counterparts of the $S^2(N)$, i.e., suitable generalizations for the $K<0$ JP vortices.

\section{Imbedding into conformally flat spacetimes}
\label{sec7:associated_spactimes}

Let us now consider the simplest relativistic spacetime that can be obtained by imbedding $S^2(N)$ into it. This is, clearly, $S^2 (N) \times \mathbb{R}$, the three-dimensional space with metric
\begin{equation}
\label{LE:metric_ansatz}
g^{(3)}_{\mu\nu} = \begin{pmatrix}
1     & 0\\
0 & -g^{(2)}_{\alpha\beta}
\end{pmatrix}.
\end{equation}
with $\mu, \nu = 0,1,2$, and $\alpha, \beta = 1,2$, where the spatial line element, $dl^2 = g^{(2)}_{\alpha\beta} d x^{\alpha} d x^{\beta}$, is that of $S^2 (N)$.

This is an interesting problem, on its own mathematical right, but it could also be of practical use when the two-dimensional membrane is made of a Dirac material, see \cite{Iorio_weyl_symmetry}. In that case the metric (\ref{LE:metric_ansatz}) is the one experienced by the low-energy massless Dirac quasiparticles of that material, realizing interesting instances of quantum field theories over nontrivial backgrounds. Examples have been proposed in \cite{ioriolambiase}, and put into contact with surfaces of constant negative $K$ (Beltrami surface, for the Rindler spacetime, hyperbolic pseudosphere for black-hole, and elliptic pseudosphere for de Sitter). Further research has been carried out on the theoretical side, see, e.g., \cite{ioriopaisvaria}, and the recent \cite{ioriopaistorsion}, as well as on the side next to experiments, see, e.g., \cite{grapheneExperiments2},\cite{grapheneExperiments}.

For what matters this paper, it is important to recall that, when $K$ is constant, positive or negative, the metric in (\ref{LE:metric_ansatz}) is necessarily conformally flat, hence the Weyl invariant Dirac theory living on it, enjoys particularly symmetric settings that allows for exact solutions, and highly improve the study. On this see \cite{Iorio_weyl_symmetry}, and also the review \cite{Iorio_curved_spacetime}.

Let us then proceed to study $S^2 (N) \times \mathbb{R}$, starting by writing the spacetime interval in isothermal spatial coordinates
\begin{equation}
\label{eq:spacetime_interval_JP}
ds_V^2 = dt^2 - \frac{4N^2}{K}\frac{\tilde{r}^{2(N-1)}}{(1+\tilde{r}^2)^2} \, d\tilde{r}^2 - \frac{4N^2}{K}\frac{\tilde{r}^{2N}}{(1+\tilde{r}^2)^2} \, d\tilde{\theta}^2 \,,
\end{equation}
where $V$ stands for ``vortices''.

By recalling that $[R_+'(\tilde{r})]^2 = \left[\frac{2N^2}{\sqrt{K}}\frac{\tilde{r}^{N-1}(1-\tilde{r}^{2N})}{(1+\tilde{r}^{2N})^2}\right]^2$, we can rewrite the above as
\begin{equation}
\label{eq:spacetime_interval_similar_to_LTB}
ds_V^2 = dt^2 - \frac{[R_+'(\tilde{r})]^2}{1 +2E(\tilde{r}) } \left( \frac{d\tilde{r}}{N}\right)^2 - R_+^2(\tilde{r}) d\tilde{\theta}^2 \,,
\end{equation}
where
\begin{equation}
	E(\tilde{r}) \equiv -2\frac{ {\tilde{r}}^{2N}}{1 + {\tilde{r}}^{2N}} + 2\left( \frac{ \tilde{r}^{2N}}{1 + \tilde{r}^{2N}}\right)^2.
\end{equation}
Notice that the function $E(\tilde{r})$ has one global minimum, $E(\tilde{r} = 1) = -1/2$.

This form of the metric is interesting, due to its similarity to the Lema\^{i}tre-Tolman-Bondi (LTB) metric, that is the spherically-symmetric dust solution of Einstein field equations \cite{Exact_spacetime}
\begin{equation}
\label{VS:eq:Lemaitre_Tolman_Bondi_metric}
ds^2 = dt^2 - \frac{(\mathcal{R}')^2}{1 + 2E} dr^2 - \mathcal{R}^2 d\Omega^2 \,,
\end{equation}
where $d\Omega^2 = d\theta^2 + \sin^2\theta d\nu^2$, and the coordinates $r$, $\theta$ and $\nu$ are standard radial distance and spherical angles, respectively (we are in four dimensions there, but it is easy to move to three by simply dropping $\nu$). Furthermore, $\mathcal{R} = \mathcal{R}(t,r)>0$ is the time-dependent radial function, and $\mathcal R'\equiv{d\mathcal{R}}/{dr}$. Finally, to avoid the singularity, it is assumed that $E = E(r) > -1/2$.

Moreover, $\mathcal{R}$ there satisfies
\begin{equation}
\label{eq:LTB_condition}
\dot{\mathcal{R}}^2 = \frac{2M}{\mathcal{R}} + 2E,
\end{equation}
where $\dot{\mathcal{R}}\equiv{d\mathcal{R}}/{dt}$, $M(r)$ is the total invariant mass within a shell of radius $r$ and $E(r)$ is the energy per unit mass
 at the radius $r$. In the static case, that is surely the case of (\ref{LE:metric_ansatz}), we have $\dot{\mathcal{R}} = 0$, and from (\ref{eq:LTB_condition}) we obtain
\begin{equation}
\label{eq:static_mass}
	M = -E \mathcal{R} \,.
\end{equation}
Thus (\ref{VS:eq:Lemaitre_Tolman_Bondi_metric}) becomes
\begin{equation}
\label{VS:eq:Lemaitre_Tolman_Bondi_metric2}
ds^2 = dt^2 - \frac{1}{1 -\frac{2M}{\mathcal{R}}} d\mathcal{R}^2 - \mathcal{R}^2 d\Omega^2 \,.
\end{equation}
We now come back to our spacetime (\ref{eq:spacetime_interval_similar_to_LTB}), and use (\ref{VS:eq:radial_part}) to write $\tilde{r} (R_+)$
\begin{equation}
\label{eq:transformation_q_to_R}
\tilde{r}^N = \frac{\frac{2N}{\sqrt{K}} \pm \sqrt{\frac{4N^2}{K} - 4R_+^2} }{2R_+} \,,
\end{equation}
where the two signs come from having to solve a square root. This means that we have to consider upper ($\tilde{r}\in (1,\infty]$) and lower ($\tilde{r}\in [0,1)$) part of $S^2(N)$,
with the locus of points $\tilde{r} = 1$ being the singular equator. We are free to choose the lower sector to correspond to the minus sign. Then the expression for $E(R_+)$ gets simplified to
\begin{equation}
\label{eq:FLRW:energy}
	E(R_+) = -\frac{K}{2N^2}R_+^2 \,,
\end{equation}
with which\footnote{We could have started with coordinates $(u,\tilde{\theta})$, instead of $(\tilde{r},\tilde{\theta})$, and with
\[
	ds_V^2 = dt^2 - du^2 - R^2_+(u)d\tilde{\theta}^2 \,.
\]
Of course, the change of variables from $u \rightarrow R_+$ leads to the same result (\ref{eq:FLRW_our_study}).}
\begin{equation}
\label{eq:FLRW_our_study}
	ds_V^2 = dt^2 - \frac{dR_+^2}{1-\frac{K}{N^2}R_+^2} \frac{1}{N^2} - R_+^2d\tilde{\theta}^2.
\end{equation}
This is a valuable result, because the metric is similar to the Friedmann--Lema\^{i}tre--Robertson--Walker (FLRW) metric, that in cosmology describes a homogeneous and isotropic universe \cite{Exact_spacetime}
\begin{equation}
\label{eq:FLRW}
ds^2 = dt^2 - \frac{ dr^2}{1-\frac{k}{A^2}r^2} - r^2d\tilde{\Omega}^2 \,,
\end{equation}
where $A$ is here the time independent scale factor, and $k/A>0$ is the (total) Gaussian curvature. Except for the factor $1/N^2$ in (\ref{eq:FLRW_our_study}), the two metrics are formally the same and, for $N = 1$, they coincide. In this latter case, $N=1$, the way $E$ and $R_+$  are related in  (\ref{eq:FLRW:energy}), tells that we deal with a situation encountered in homogeneous, isotropic, and (since the total positive Gaussian curvature is such that $k/A^2 \equiv {K}/{N^2} = 1/(aN)^2$, also) finite universe, described by a FLRW metric. This is the simplest Einstein static universe what we may deal with.

\section{Conclusions}

We have found the two-dimensional surfaces corresponding to the JP non-topological vortex solutions discussed in \cite{Horvathy_Yera}. They are the bulge surfaces of constant positive Gaussian curvature, $K=1/a^2$, characterized by an integer $N$ (the vortex index), related to the surface's maximal radius through $c = a N$. We called them $S^2 (N)$. Noticeably, the range of the latitude gives the full bulge surface, not just a portion. As for the longitude, we choose solutions for which the full $2\pi$ angle is obtained. The sphere is only one of the infinitely many cases, and corresponds to $N=1$.

One might think of using graphene, or any other Dirac material, to construct $S^2 (N)$s in a laboratory. In that case our findings could be of interest to realize table-top Dirac massless excitations on such nontrivial backgrounds. To this end, it would be interesting to study the effects on our results of the findings of \cite{Horvathytopology}, on the topology of these solutions, as well as of the findings of  \cite{Horvathyspinor}, where the generalization to spinors is considered.

Furthermore, we have also briefly addressed the question of what kind of spacetimes of the form $S^2 (N) \times \mathbb{R}$, can be associated to these surfaces. Naturally, LTB and FLRW spacetimes emerge. The application of this to analog realizations in the laboratory, by using Dirac materials, needs further study. First, one should fully take control of the effects of the edges, then the physical meaning of the gravitational quantities need be understood in terms of their condensed matter counterparts.

\section*{Acknowledgments}

A.~I. is partially supported by the grant UNCE/SCI/013.

\appendix

\section{An alternative to bulge surfaces as JP vortices}

Let us write (\ref{eq:line_element_generally}) in the following form
\begin{eqnarray}
	dl^2  &=& \frac{4N^2}{K}\frac{\tilde{r}^{2(N-1)}}{(1+\tilde{r}^{2N})^2}(d\tilde{r}^2+\tilde{r}^2d\tilde{\theta}^2) \nonumber \\
          &=& \frac{1}{K}\frac{N^2}{\left( \frac{\tilde{r}^{N}+\tilde{r}^{-N}}{2}\right)^2}\left[ \left( \frac{d\tilde{r}}{\tilde{r}}\right)^2 + d\tilde{\theta}^2\right] = \frac{a^2}{  \cosh^2\tilde{R}} \left(d\tilde{R}^2 + d\tilde{\omega}^2\right)
\end{eqnarray}
where in the last equality we used $\tilde{R}\equiv\ln \tilde{r}^{N}$, $\tilde{\omega} \equiv N\tilde{\theta}$ and, as usual, $K = {1}/{a^2}$. In \cite{Iorio_curved_spacetime} it was shown that this line element corresponds to surfaces of revolution with constant $K>0$. Indeed, with \cite{Iorio_curved_spacetime}
\begin{equation}
\tilde{\omega} \equiv v,~\tilde{R} \equiv \ln\left(1+\frac{2}{\cot(u/2a)-1} \right),
\end{equation}
one obtains
\begin{equation}
dl^2 = \frac{a^2}{  \cosh^2\tilde{R}} \left(d\tilde{R}^2 + d\tilde{\omega}^2\right) = du^2 + a^2\cos^2\frac{u}{a}dv^2 \,,
\end{equation}
therefore, clearly, $R(u) = a\cos({u}/{a})$, and we have a sphere. In the isothermal coordinates
\begin{equation}
\label{eq:radial_function}
R(\tilde{r})  \equiv R(u(\tilde{R}(\tilde{r}))) =  \frac{2}{\sqrt{K}}\frac{\tilde{r}^N}{\tilde{r}^{2N} + 1} \,.
\end{equation}
With the canonical parametrization (\ref{eq:surfaces_of_revolution_parametrization}) we can write
\begin{equation}
\label{VS:eq:coordinates_xy_sphere1}
x(\tilde{r},\tilde{\theta}) = \frac{2}{\sqrt{K}}\frac{\tilde{r}^N}{\tilde{r}^{2N} + 1}\cos(N\tilde{\theta}),~~
y(\tilde{r},\tilde{\theta}) = \frac{2}{\sqrt{K}}\frac{\tilde{r}^N}{\tilde{r}^{2N} + 1}\sin(N\tilde{\theta}).
\end{equation}
Since the surfaces of our interest here have circular symmetry, this parametrization is supported. In fact, there is one obscure point with (\ref{VS:eq:coordinates_xy_sphere1}), related to the $N$-fold rotation around the $z$ axis. We shall get back to it later.

It is also easy to check that the parametrization satisfies Eqs. (\ref{eq:coor_transf_2}) and (\ref{eq:coor_transf_3}). Hence, the $z$ coordinate cannot depend on $\tilde{\theta}$. Having $x,y$ coordinates, the $z$ coordinate can be obtained from (\ref{eq:coor_transf_1})
\begin{equation}
\label{VS:eq:coordinates_z_sphere1_param}
z(\tilde{r}) = \int \sqrt{\phi_{+}^2(\tilde{r}) - [R(\tilde{r})']^2} d\tilde{r} = -\frac{2}{\sqrt{K}}\frac{1}{1+\tilde{r}^{2N}} \,.
\end{equation}
It is now straightforward to show that the expression $\phi_{+}^2(\tilde{r}) - [R(\tilde{r})']^2$ is non-negative for all $\tilde{r}\in [0,+\infty]$, hence the $z$ coordinate enjoys this utmost
range.

We may also define new radial and angular coordinates: $\eta \equiv \tilde{r}^N$ and $\tilde{\omega} \equiv N\tilde{\theta}$,  respectively. Then the spatial coordinates are:
\begin{equation}
\label{VS:eq:coordinates_xyz_sphere_1}
x(\eta,\tilde{\omega}) = \frac{2}{\sqrt{K}}\frac{\eta}{\eta^2 + 1}\cos\tilde{\omega},~
y(\eta,\tilde{\omega}) = \frac{2}{\sqrt{K}}\frac{\eta}{\eta^2 + 1}\sin{\tilde{\omega}},~z(\eta) =-\frac{2}{\sqrt{K}}\frac{1}{1+\eta^{2}},
\end{equation}
where $\eta \in [0,+\infty]$ and $\tilde{\omega} \in [0, 2\pi N]$.

We may modify (\ref{VS:eq:coordinates_xyz_sphere_1}) by a following substitution: $\frac{\eta}{\eta^2 + 1} \equiv \frac{1}{2}\sin \tilde{\phi}$, where $\tilde{\phi}\in[0,\pi]$. For a choice $\tilde{\phi} = 0 \Leftrightarrow \eta=0$ and  $\tilde{\phi} = \pi \Leftrightarrow \eta=+\infty$ we get $\cos\tilde{\phi} = - \frac{\eta^2 - 1}{\eta^2 + 1}$, so $ -\frac{1}{\eta^2+1} =  - \frac{1}{2}\cos\tilde{\phi} - \frac{1}{2} $.  Finally, assuming $1/\sqrt{K}\equiv a$, we may rewrite the formulae for the spatial coordinates as
\begin{equation}
\label{VS:eq:coordinates_xyz_sphere_1b}
x = a\sin{\tilde{\phi}}\cos {\tilde{\omega}},
~y = a\sin {\tilde{\phi}}\sin {\tilde{\omega}},
~z = -a\cos {\tilde{\phi}} - a,
\end{equation}
where $\tilde{\phi}\in[0,\pi]$ and $\tilde{\omega}\in[0,2\pi N]$.

This is a full sphere, for any arbitrary $N$. It is clearly a solution, but then the role of the $N$-folding needs be clarified from a physical point of view.

\section{Proof of a useful identity}
Here we recall and prove the identity (\ref{eq:useful_identity})
\begin{equation}
2 \arctan\sqrt{p} = \arcsin\left( \frac{p-1}{p+1}\right) +\frac{\pi}{2}\,.
\end{equation}
We define a new function
\begin{equation}
	f(p) \equiv 2 \arctan\sqrt{p} -  \arcsin\left( \frac{p-1}{p+1}\right) + c\,,
\end{equation}
where $c$ is an unknown constant,
and show that its first derivative is zero
\begin{equation}
	\frac{d}{d p} f(p) = \frac{1}{\sqrt{p}(1+p)} - \frac{1}{\sqrt{1-\left( \frac{p-1}{p+1}\right)^2}} \frac{2}{(1+p)^2} 	= \frac{1}{\sqrt{p}(1+p)} - \frac{1}{\sqrt{4p}}\frac{2}{1+p} = 0\,.
\end{equation}
Once $f$ is a constant function, we want to find the appropriate constant $c$ to prove (\ref{eq:useful_identity}), so $f(p)$ = 0. For any suitable value of $p$, say $p = 1$, we obtain
\begin{equation}
	f(1) = \frac{\pi}{2} + c = 0\,,
\end{equation}
therefore $c = -{\pi}/{2}$.

%
%
%
%
%

\end{document}